\documentclass[aps,showpacs,showkeys,floatfix,amsmath]{revtex4}
\usepackage{graphicx} 
\usepackage{dcolumn} 

\begin{document} 
 
\title{Low densities in asymmetric nuclear matter}
 
\author{J\'er\^ome Margueron} 
\affiliation{Institut de Physique Nucl\'eaire, Universit\'e 
Paris-Sud, IN$_2$P$_3$-CNRS, F-91406 Orsay Cedex, France}
\author{Eric van Dalen} 
\affiliation{Departament d'Estructura i Constituents de la Mat\`eria,
Universitat de Barcelona, Diagonal 647, 08028 Barcelona, Spain} 
\author{Christian Fuchs} 
\affiliation{Institut f\"ur Theoretische Physik,  
Universit\"at T\"ubingen, D-72076 T\"ubingen, Germany}

\date{\today} 
 
\begin{abstract} 
Asymmetric nuclear matter is investigated in the low density region below 
the nuclear saturation density. Microscopic calculations based on the 
Dirac Brueckner Hartree-Fock (DBHF) approach with realistic 
nucleon-nucleon potentials are used to adjust a low density functional.
This functional is constructed on a density expansion of the
relativistic mean field theory which allows a clear interpretation of the
role of the mesons to the equation of state.
It is shown that a correction term should be added to the functional in
order to take into account the effects beyond the mean field.
Two functionals with different corrections are obtained and their
topological properties has been studied.
Those functionals converge to predict a reduction of the spinodal zone
in asymmetric nuclear matter by about 15-20\% and an isoscalar
unstable mode closer to the constant Z/A direction than the functional
without correction.
\end{abstract} 
 


\maketitle 

How does nuclear matter properties change when the density decreases
from saturation densities ? In this low density regime, what is the
role of the isospin asymmetry ?
Indeed, the nuclear density functional below saturation density has
not been much studied while it has an importance for several topics
concerning atomic nuclei surface properties, the equation of state of
the core of neutron stars and for the dynamical description of heavy
ion collision, both at intermediate and relativistic energies.  
Recently, several attempts have tried to establish a relation between
the low density equation of state and the nuclear properties like
surface behavior and pairing properties~
\cite{bal04,Sedrakian:2006mq,Sedrakian:2003cc,Gogelein:2007pb}, neutron 
radii~\cite{bro00} or the spinodal instability~\cite{bar06}.
Those works are based on phenomenological density functionals or fits
of ab-initio calculations but without considering the very low
properties of the equation of state.
Below saturation density, effects of two-body correlations are
important and induce anomalies in the density dependence of the
equation of state \cite{horowitz05}.

Models which make predictions for the nuclear equation of state (EOS) can roughly be divided  
into three classes: phenomenological density functionals, effective field 
theory approaches,
and ab initio approaches. The phenomenological density functionals 
are based on effective density dependent interactions such as  
Gogny or Skyrme forces \cite{reinhard04,gogny} or  
relativistic mean field (RMF) models \cite{rmf}. Parameters are adjusted to   
nuclear bulk properties and finite nuclei. 
Effective field theory (EFT) approaches are based on a perturbative
expansion of the nucleon-nucleon 
interaction or the nuclear mean field within power counting
schemes. These approaches lead to a more systematic expansion of the  
EOS in powers of density, respectively the Fermi momentum $k_F$.
The EFT approaches can be based on density functional 
theory~\cite{eft1,eft2} or e.g. on chiral perturbation 
theory~\cite{lutz99,finelli}. 
The advantage of EFT is the small number of free parameters and a
correspondingly higher predictive power.
However, when high precision fits are intended, the EFT functionals are  
based on approximately the same number of model parameters as   
phenomenological density functionals due to 
fine tuning through additional parameters.

Ab initio approaches are based on high precision  
free space nucleon-nucleon interactions. 
In addition, predictions for the nuclear EOS  
are parameter free. Examples of such approaches are variational  
calculations \cite{wiringa79,akmal98}, Brueckner-Hartree-Fock (BHF)  
\cite{jaminon,zuo02,zuo04,baldo07}  
or relativistic Dirac-Brueckner-Hartree-Fock (DBHF)  
\cite{terhaar87,bm90,dejong98,boelting99,honnef,dal04}  
calculations and Green function Monte-Carlo approaches (GFMC) 
\cite{carlson03,dickhoff04,Fabrocini:2006xt}. Non-relativistic ab
initio calculations do not meet the empirical region of saturation,
whereas relativistic calculations do a better job. 
This deficiency can be solved by the explicit inclusion 
of three-body forces where the relativistic approach accounts 
already effectively for part of these contributions. For a more 
detailed discussion see e.g. Ref.~\cite{baldo07}.

In the present work microscopic calculations based on the DBHF 
approach~\cite{dal04} using a realistic nucleon-nucleon
potential, i.e. the Bonn A interaction \cite{bonn},  
are used to obtain a functional which describes the equation of
state from low densities up to saturation density. 
The construction of this functional is motivated by relativistic mean field (RMF) 
theory~\cite{sw86,ring96}. Since both, DBHF and RMF are relativistic approaches 
based Dirac phenomenology they have similar features, in particular the 
occurrence of large and cancellating scalar and vector fields in the 
isoscalar sector. That the occurrence of such fields is a fundamental consequence 
of the elementary nuclear force has recently been 
shown in \cite{plohl06}. Similar fields, which are, however, smaller, occur 
also in the isovector sector. Thus, RMF theory is well suited for the 
present investigations. 
However, RMF theory is insufficient to reproduce the 
more complex nonlinear behavior of the DBHF energy density near $\rho_B=0.1$ fm$^{-3}$, 
where for instance effects of the deuteron pole show up.
Corrections beyond mean field are necessary and the non-linear
behavior is then corrected in adding new terms in the functional. 

The paper is organized as follows: the relativistic DBHF
is shortly sketched in Sec.~\ref{sec:RBA}. 
Furthermore, Sec.~\ref{sec:sebd} is devoted to the relativistic
mean-field approach and the series expansion in the baryonic density. 
This expansion allows a clear interpretation of the meson contributions
to the equation of state.
The parameters of the functionals are obtained in Sec.~\ref{sec:param}.
With these density functionals the dynamics of the liquid-gas 
transition induced by heavy ion collision at Fermi energies, i.e. the
spinodal instabilities, are investigated and analyzed in Sec.~\ref{sec:si}.
Finally, we end with a conclusion in Sec.~\ref{sec:c}.

\section{DBHF approach} 
\label{sec:RBA}
We consider homogeneous nuclear matter at low density so that 
two-body correlations dominate. Of course, at very low density, clustering 
phenomena can occur, like deuteron, tritium, helium and alpha particle 
formation \cite{horowitz05,Tohsaki:2001an,Beyer:2000ds}. 
Therefore, we are going to consider densities which are low with 
respect to saturation density, but still large compared to typical 
values where the onset of clustering occurs. 
Typically, we are considering densities between about one tenth to 
one half of the saturation density. In this density region, microscopic 
calculations based on the relativistic Dirac-Brueckner-Hartree-Fock (DBHF) 
approach \cite{terhaar87,honnef,boelting99,dal04,dal05}  are expected
to be quite accurate, and they will be, therefore, the starting point
of our analysis.  
\begin{figure}[htb] 
\center 
\includegraphics[scale=0.46]{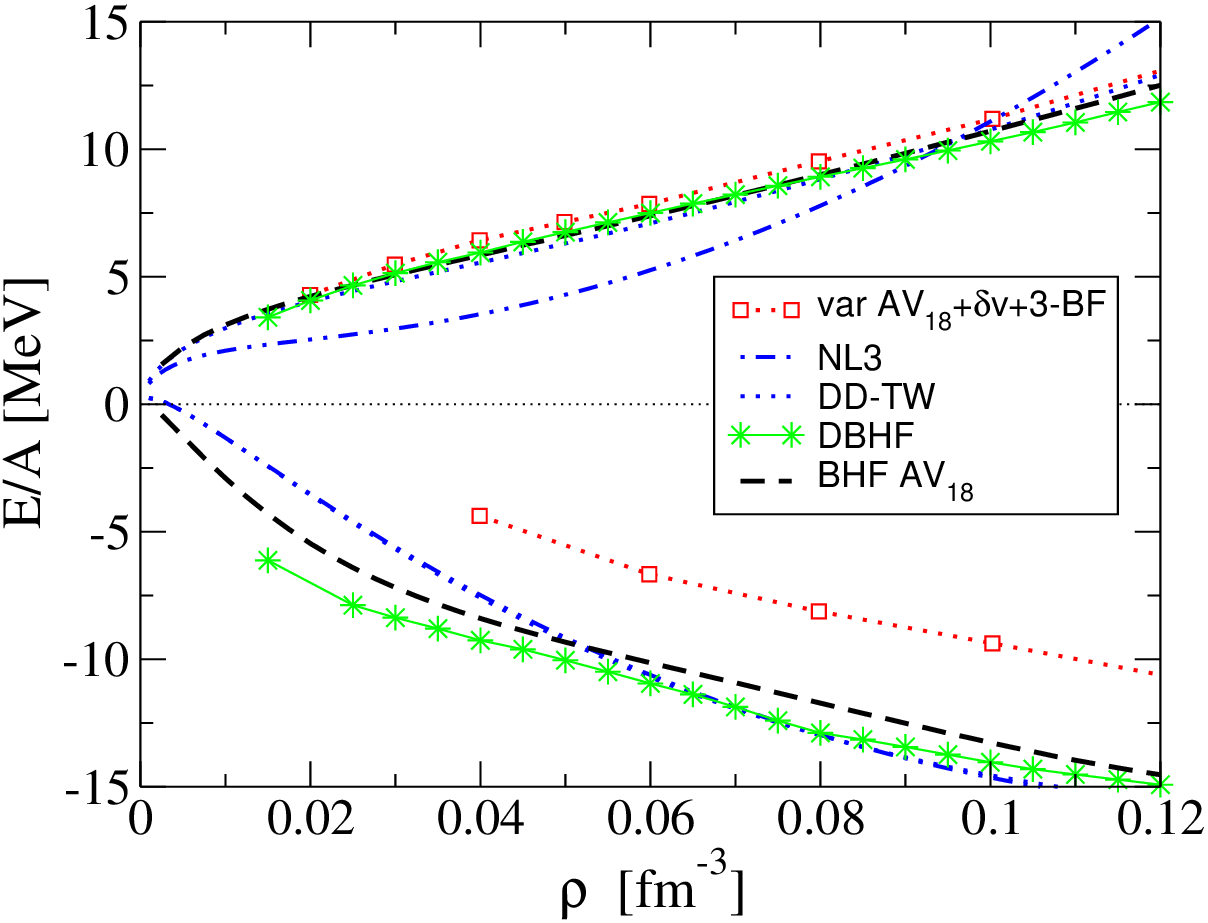} 
\caption{(Color online) Comparison of several equation of state with the DBHF results. 
The DBHF calculation are represented by the stars, the variational by the empty 
squares, the non-relativistic Brueckner results by the dashed line and
two different relativistic mean field parametrisations: NL3
(dashed-dotted line) and DD-TW (dotted line). } 
\label{fig01} 
\end{figure} 
 
In the relativistic Brueckner approach the nucleon  
inside the medium is dressed by the self-energy $\Sigma$.  
The in-medium T-matrix which is obtained from  
the relativistic Bethe-Salpeter (BS) equation plays the role  
of an effective two-body interaction which contains all short-range  
and many-body correlations of the ladder approximation.  
Solving the BS-equation the Pauli principle is respected  
and intermediate scattering states are projected  
out of the Fermi sea.  
The summation of the  T-matrix over the occupied states inside the Fermi sea  
yields finally the self-energy in Hartree-Fock approximation. This
coupled set of equations states a self-consistency problem which has
to be solved by iteration. Technical details of the present DBHF 
calculations, in particular the treatment of isospin asymmetry, can 
be found in \cite{dal04}. The results are based on the Bonn A 
one-boson-exchange potential for the bare nucleon-nucleon interaction
\cite{bonn}.  

Fig.~(\ref{fig01}) compares the prediction for the low density regime of 
symmetric nuclear matter (below zero) and pure neutron matter 
(above zero) from microscopic many-body approaches. It shows in 
addition the equations-of-state obtained by two typical 
phenomenological relativistic mean field models. From the microscopic 
side these are the DBHF results \cite{dal04} which will be further 
analyzed, non-relativistic  Brueckner (BHF) calculations from \cite{bal04} 
and variational calculations from \cite{akmal98}. The variational and 
BHF calculations are based on the ${\rm AV}_{18}$ Argonne  
potential. The variational calculations include in addition 
phenomenological 3-body-forces (Urbana IX) and relativistic boost 
corrections which both do, however, not play an important role 
in the low density regime. The phenomenological models are the 
well established NL3 \cite{lal97} relativistic mean field (RMF) 
parametrisation and the RMF model of \cite{typel99} (DD-TW). The latter 
is a phenomenological version of density dependent  RMF theory 
using density dependent meson-nucleon couplings \cite{fule95} which 
allows for a larger freedom in the adjustment of the EOS. Both 
approaches fit finite nuclei properties with high accuracy. 

The first what becomes evident from Fig.~(\ref{fig01}) is the
remarkable agreement of the microscopic approaches (DBHF, BHF,
variational) concerning the pure neutron matter EOS. 
This indicates that both, the interaction 
and the many-body schemes are well controlled in the $nn$ sector 
at low densities. The reason is on the one handside the large 
$nn$ scattering length and on the other side the lack of clustering 
phenomena ($d$, $\alpha$ etc) which make the treatment of neutron matter 
at subnuclear densities less model dependent. In this context it is 
worth noticing that the microscopic calculations (BHF/DBHF, variational) are 
consistent with the low density limit of 
 'exact' Quantum-Monte-Carlo calculations for neutron matter \cite{carlson03} 
and with the result of the renormalization group $V_{\rm low k}$ 
approach \cite{lowk03}. 

The situation seems to be different for symmetric 
nuclear matter. The Brueckner calculations show significantly more 
binding than the variational calculations of \cite{akmal98}. On the other 
hand, the DBHF and BHF results are very close and exhibit the same 
low density behavior: in contrast to RMF theory and also to \cite{akmal98} 
one can observe a non-linear convergence to zero when 
the density decreases. This fact is associated to the deuteron channel and 
can possibly be interpreted as a manifestation of the onset of the 
super-fluid phase. In this context it is interesting to note that 
a recent study of low density nuclear matter \cite{horowitz05}, based on 
a virial expansion which includes protons, neutrons and $\alpha$-particle 
degrees of freedom, revealed a low density EOS which is in qualitative 
agreement with the DBHF predictions. In the virial approach the binding 
energy goes smoothly to zero for neutron matter while the energy per 
particle $E/A$ minus the (free) kinetic energy in symmetric matter 
remains practically constant at a value around -8 MeV down to extremely 
low densities ($\rho_B \simeq 0.0002~{\rm fm}^{-3}$) before it rapidly 
drops to zero (Fig. 15 in \cite{horowitz05}). 
Subtracting from the DBHF result the kinetic energy of 
a non-relativistic Fermi gas $3k_{F}^2/10M$ yields at 
$k_F = 0.5~{\rm fm}^{-1}$ ($\rho_B = 0.0084~{\rm fm}^{-3}$) a values 
of -8.4~MeV which coincides remarkably well with the virial low density 
limit.

\section{Series expansion of the RMF Lagrangian in the baryonic density} 
\label{sec:sebd}

A Lagrangian density of interacting many-particle 
system consisting of nucleons, isoscalar (scalar $\sigma$, vector $\omega$), 
and isovector (scalar $\delta$, vector $\rho$) mesons is the starting 
point of the relativistic mean field (RMF) theory,
\begin{widetext} 
\begin{eqnarray}\label{eq:1} 
{\cal L } &=& \bar{\psi}[i\gamma_{\mu}\partial^{\mu}-(M- 
g_{\sigma}\sigma -g_{\delta}\vec{\tau}\cdot\vec{\delta}) 
-g{_\omega}\gamma_\mu\omega^{\mu}-g_\rho\gamma^{\mu}\vec\tau\cdot 
\vec{\rho}_{\mu}]\psi \nonumber \\&& 
+\frac{1}{2}(\partial_{\mu}\sigma\partial^{\mu}\sigma-m_{\sigma}^2\sigma^2) 
-U(\sigma)+\frac{1}{2}m^2_{\omega}\omega_{\mu} \omega^{\mu} 
+\frac{1}{2}m^2_{\rho}\vec{\rho}_{\mu}\cdot\vec{\rho}^{\mu} \nonumber 
\\&& 
+\frac{1}{2}(\partial_{\mu}\vec{\delta}\cdot\partial^{\mu}\vec{\delta} 
-m_{\delta}^2\vec{\delta^2}) -\frac{1}{4}F_{\mu\nu}F^{\mu\nu} 
-\frac{1}{4}\vec{G}_{\mu\nu}\vec{G}^{\mu\nu}, 
\end{eqnarray} 
\end{widetext} 
where $\sigma$ is the $\sigma$-meson field, 
$\omega_{\mu}$ is the $\omega$-meson field, $\vec{\rho}_{\mu}$ is 
$\rho$ meson field, $\vec{\delta}$ is the isovector scalar field of the 
$\delta$-meson. 
$F_{\mu\nu}\equiv\partial_{\mu}\omega_{\nu}-\partial_{\nu}\omega_{\mu}$, 
$\vec{G}_{\mu\nu}\equiv\partial_{\mu}\vec{\rho}_{\nu}-\partial_{\nu}\vec{\rho}_{\mu}$, 
and the $U(\sigma)$ is a nonlinear potential of $\sigma$ meson : 
$U(\sigma)=\frac{1}{3}a\sigma^{3}+\frac{1}{4}b\sigma^{4}$. 
Dynamical equations deduced at the mean field approximation are
presented in the appendix~\ref{app:rmfm}. We refer to the appendix
for all the standard definitions. Hereafter, we introduce the coupling
constants  
$f_i=g_\rho/m_i$ for $i$=$\sigma$, $\delta$, $\omega$ and $\rho$, 
and the non-linear constant $f_\sigma^{nl}=a (f_\sigma/m_\sigma)^3$. 
In the following the energy density $\epsilon$ is expanded
up to the power 4 in proton and neutron densities.
The expression of the density of energy is given in the
appendix~\ref{app:rmfm}. 
For the linear version of the RMF model, i.e. without a nonlinear 
$\sigma$-meson potential $U(\sigma)$ such an expansion can be found 
in \cite{serot97}. Here we extend this expansion to the non-linear 
case and to the isospin sector, i.e. to $\rho_p \neq \rho_n$ 
(the isovector mesons $\rho$ and $\delta$ are included).
Notice however, that in the present form, only the non-linear term 
with the coupling constant $a$ is included because the one with the
coupling constant $b$ contributes to higher terms in the density 
expansion.

The scalar field is the solution of the following self-consistent equation
\begin{eqnarray} 
g_\sigma \sigma = f_\sigma^2 \rho_s - \frac{a}{g_\sigma m_\sigma^2}
(g_\sigma\sigma)^2 - \frac{b}{(m_\sigma g_\sigma)^2}(g_\sigma\sigma)^3~. 
\label{scalarself}
\end{eqnarray} 
A low density approximate solution is presented in the appendix~\ref{app:nls}. 
The solution is expressed as a function of the scalar density
$\rho_s$, 
\begin{eqnarray} 
g_\sigma \sigma &=& f_\sigma^2 \rho_s - f_\sigma^{nl} \rho_s^2+ { o}(\rho_{si}^3)~ . 
\label{scalar}
\end{eqnarray} 
The low-density expansion of the scalar density $\rho_{si}$ is then
required. It yields  
\begin{eqnarray} 
\rho_{si}=\rho_i- 
\frac{3}{10M_i^{*2}}\left(\frac{6\pi^2}{\gamma}\right)^{2/3}\rho_{i}^{5/3}+
\frac{9}{56M_i^{*4}}\left(\frac{6\pi^2}{\gamma}\right)^{4/3}\rho_{i}^{7/3}-
\frac{15}{144M_i^{*6}}\left(\frac{6\pi^2}{\gamma}\right)^2\rho_{i}^3+
\frac{105}{1408M_i^{*8}}\left(\frac{6\pi^2}{\gamma}\right)^{8/3}\rho_{i}^{11/3}
+{ o}(\rho_i^4)~, \nonumber\\
\label{rhos}
\end{eqnarray} 
where $\gamma$ is the degeneracy of the system.
Note that this expansion is also a relativistic expansion in the
parameter $k_{Fi}/M^*_i$. 

Neutron and proton Dirac masses, also called the scalar masses, 
are expressed in terms of the
scalar and isoscalar fields. Using Eq.~(\ref{scalar}) and Eq.~(\ref{rhos}),
The low density expansion of the Dirac masses is given by 
(-~proton, +~neutron), 
\begin{eqnarray} 
M_i^*=M&-& 
f_\sigma^2 \left[\rho_B 
-\frac{3}{10M^2}(3\pi^2)^{2/3}\left(\rho_p^{5/3}+\rho_n^{5/3}\right) 
+\frac{9}{56M^4}(3\pi^2)^{4/3}\left(\rho_p^{7/3}+\rho_n^{7/3}\right) 
-\frac{3}{5M^3}f_\sigma^2(3\pi^2)^{2/3}\left(\rho_p^{8/3}+\rho_n^{8/3}\right) 
\right]\nonumber \\ 
&\pm& 
f_\delta^2 \left[\rho_3 
-\frac{3}{10M^2}(3\pi^2)^{2/3}\left(\rho_p^{5/3}-\rho_n^{5/3}\right) 
+\frac{9}{56M^4}(3\pi^2)^{4/3}\left(\rho_p^{7/3}-\rho_n^{7/3}\right) 
-\frac{3}{5M^3}f_\delta^2(3\pi^2)^{2/3}\left(\rho_p^{8/3}-\rho_n^{8/3}\right) 
\right]\nonumber \\ 
&+&f_\sigma^{nl}\left[\rho_B^2 
-\frac{3}{5M^2}(3\pi^2)^{2/3}\rho_B\left(\rho_p^{5/3}+\rho_n^{5/3}\right)  
\right] 
+{ o}(\rho_i^3). 
\label{eq:effmass} 
\end{eqnarray}

\begin{figure}[tb] 
\center 
\includegraphics[scale=0.5]{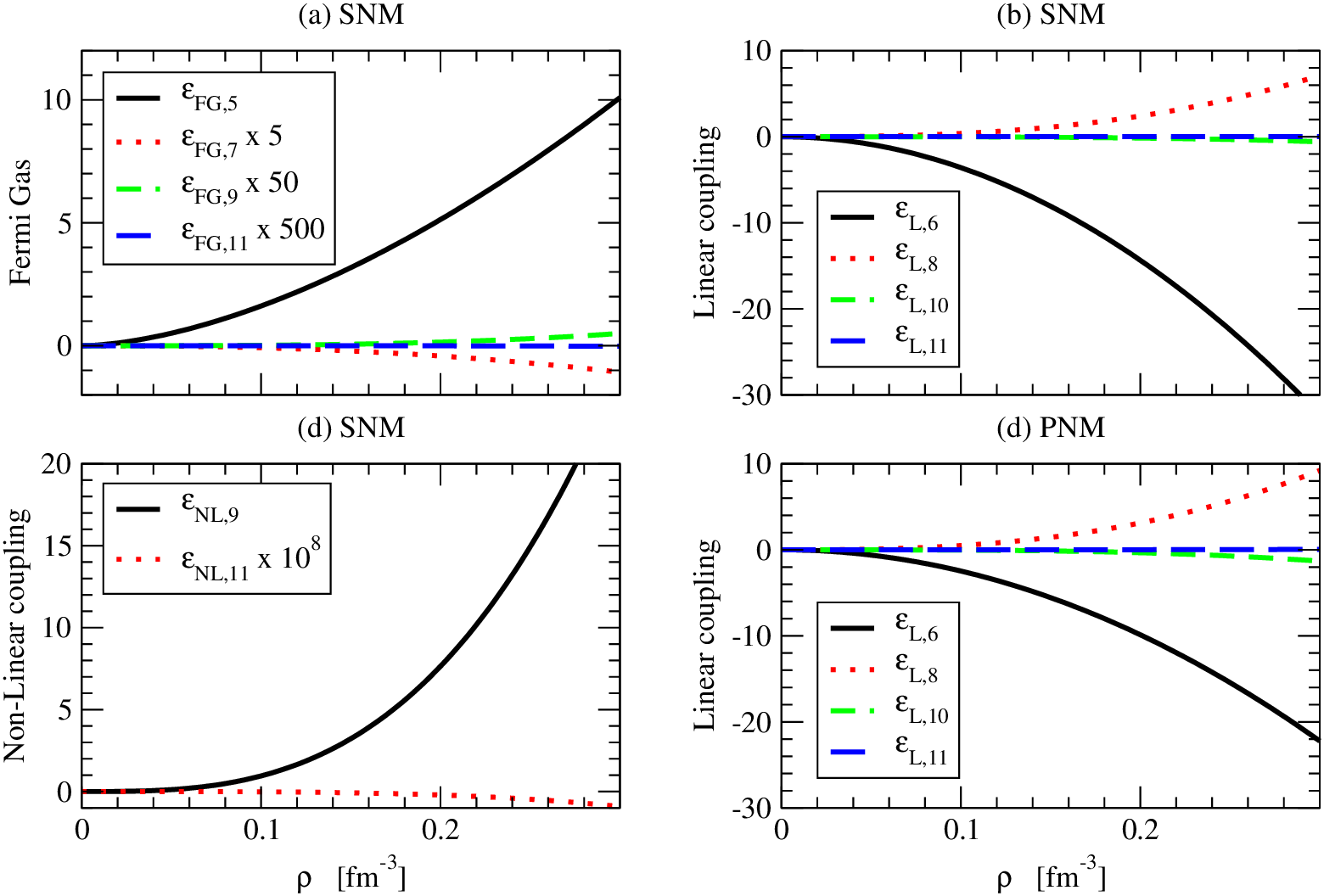} 
\caption{(Color online) Representation of the different terms in the serie expansion
of the energy functional with respect to the baryonic density, up to
0.3~fm$^{-3}$. The first 3 subsets show the fast convergence in
symmetric nuclear matter (SNM) and the last subset (d) show the
contribution of the isovector mesons in pure neutron matter (PNM).
To give an idea of the fast convergence, we have multiplied some of
the terms by huge constants.
The coupling constants (set A NL$\rho\delta$) obtained by Liu et
al.~\cite{liu05} has been used.}
\label{figu04} 
\end{figure} 
 
Now, we evaluate the full density functional,
\begin{eqnarray}
\epsilon&=&\epsilon_{kin}+\frac{1}{2}f_\sigma^2\rho_s^2
-\frac{2}{3}f_\sigma^{nl}\rho_s^2
+\frac{1}{2}f_\omega^2\rho_B^2+\frac{1}{2}f_\rho^2\rho_3^2
+\frac{1}{2}f_\delta^2\rho_{s3}^2+{ o}(\rho^4)
\end{eqnarray}
where $\epsilon{kin}$ is the kinetic term. The density functional is
decomposed into several terms:
\begin{eqnarray} 
\epsilon(\rho_n,\rho_p) =M\rho_B + \epsilon_{FG}(\rho_n,\rho_p) + 
\epsilon_L(\rho_n,\rho_p) 
+ \epsilon_{NL}(\rho_n,\rho_p) +{ o}(\rho^4) ~,
\label{edens1}
\end{eqnarray} 
where the term $\epsilon_{FG}$ is the contribution of the free Fermi
gas without the rest mass, 
the term $\epsilon_L$ is generated by the interactions and the
Dirac mass and the term $\epsilon_{NL}$ is the correction coming
from the non-linear $\sigma$ coupling. 
In the following we give explicitly the form of those terms,
classified according to the power in Fermi momentum (=power in density
divided by 3) in order to have integer index. 
The pure kinetic contributions (FG) is
$\epsilon_{FG}=\epsilon_{FG,5}+\epsilon_{FG,7}+\epsilon_{FG,9}+\epsilon_{FG,11}$
up to the power 4 in the densities where
\begin{eqnarray} 
\epsilon_{FG,5}(\rho_n,\rho_p)&=&\frac{3}{10 M}\left(3\pi^2\right)^{2/3}
\left(\rho_p^{5/3}+\rho_n^{5/3}\right) ,\\ 
\epsilon_{FG,7}(\rho_n,\rho_p)&=&-\frac{3}{56M^3}\left(3\pi^2\right)^{4/3}
\left(\rho_p^{7/3}+\rho_n^{7/3}\right) ,\\ 
\epsilon_{FG,9}(\rho_n,\rho_p)&=&\frac{1}{48M^5}\left(3\pi^2\right)^{2}
\left(\rho_p^{3}+\rho_n^{3}\right) ,\\ 
\epsilon_{FG,11}(\rho_n,\rho_p)&=&-\frac{15}{1408 M^7}
\left(3\pi^2\right)^{8/3}\left(\rho_p^{11/3}+\rho_n^{11/3}\right) . 
\end{eqnarray} 
The contribution of the mesons (with only linear couplings) and
Dirac mass contribution is 
$\epsilon_{L}=\epsilon_{L,6}+\epsilon_{L,8}+\epsilon_{L,10}+\epsilon_{L,11}$
where
\begin{eqnarray} 
\epsilon_{L,6}(\rho_n,\rho_p)&=&\frac{1}{2}\left(-f_\sigma^2+f_\omega^2 \right)\rho_B^2
+\frac{1}{2}\left(-f_\delta^2+f_\rho^2 \right)\rho_3^2 ,\\
\epsilon_{L,8}(\rho_n,\rho_p)&=&\frac{3}{10M^2}
\left(3\pi^2\right)^{2/3}\left[f_\sigma^2\rho_B 
\left(\rho_p^{5/3}+\rho_n^{5/3}\right)+f_\delta^2 \rho_3 
\left(\rho_p^{5/3}-\rho_n^{5/3}\right)\right] ,\\ 
\epsilon_{L,10}(\rho_n,\rho_p)&=&-\frac{9}{M^4}\left(3\pi^2\right)^{4/3} f_\sigma^2 
\left[\frac{1}{56}\left(\rho_p^{7/3}+\rho_n^{7/3}\right)\rho_B+ 
\frac{1}{200}\left(\rho_p^{5/3}+\rho_n^{5/3}\right)^2\right]\nonumber\\ 
&&-\frac{9}{M^4}\left(3\pi^2\right)^{4/3}f_\delta^2 
\left[\frac{1}{56}\left(\rho_p^{7/3}-\rho_n^{7/3}\right)\rho_3+ 
\frac{1}{200}\left(\rho_p^{5/3}-\rho_n^{5/3}\right)^2\right] ,\\ 
\epsilon_{L,11}(\rho_n,\rho_p)&=&\frac{3}{10M^3} \left(3\pi^2\right)^{2/3}
\left[f_\sigma^4 
\rho_B^2\left(\rho_p^{5/3}+\rho_n^{5/3}\right)+f_\delta^4
\rho_3^2\left(\rho_p^{5/3}-\rho_n^{5/3}\right) +2f_\sigma^2f_\delta^2 
\left(\rho_p^2-\rho_n^2\right)\left(\rho_p^{5/3}-\rho_n^{5/3}\right)\right] .
\end{eqnarray} 
Finally, the first order corrections induced by the non-linear
$\sigma$-coupling is $\epsilon_{NL}=\epsilon_{NL,9}+\epsilon_{L,11}$ where
\begin{eqnarray} 
\epsilon_{NL,9}(\rho_n,\rho_p) &=& \frac{1}{3}f_\sigma^{nl}\rho_B^3 ,
\label{eq:nl}\\
\epsilon_{NL,11}(\rho_n,\rho_p) &=&-\frac{3}{10M^2}f_\sigma^{nl}
\left(3\pi^2\right)^{2/3} \rho_B^2\left(\rho_p^{5/3}+\rho_n^{5/3}\right) .
\label{u1}
\end{eqnarray} 

The convergence of this series expansion is checked in symmetric
nuclear matter (SNM) and pure neutron matter (PNM) using the set of
coupling constants, set A NL$\rho\delta$, obtained by Liu et
al.~\cite{liu05}. 
We choose this set of parameters because it has been obtained with the
same degrees of freedom as the one we consider in our Lagrangian.
We show in Fig.~\ref{figu04} the contribution of the different terms
of the functional up to the power 4 in the densities. Some terms has
been multiplied by a huge factor to distinguish from each others.
The convergence is essentially due to the shorting in power of $k_F/M$
which comes with our expansion.
This figure shows how negligeable are the terms with large power
counting in the density expansion, even for the higher densities
represented (about 0.3~fm$^{-3}$).

In the following the low density RMF functional defined in 
Eqs.~(\ref{edens1})-(\ref{u1}) will be used to fit the result of the
DBHF calculation~\cite{dal04} and to calculate the spinodal
instabilities in low density asymmetric nuclear matter.

\section{Determination of the parameters of the functional
\label{sec:param}}

In this section, we fit the result of the DBHF calculation (Dirac
mass, energy density and binding energies) in the density
region between 0.01 and 0.2 fm$^{-3}$ using the low density RMF
functional defined in Eqs.~(\ref{edens1})-(\ref{u1}). 
The fits are based on the density 
expansion of the RMF Lagrangian (\ref{eq:1}) which contains non-linear
terms in the scalar $\sigma$ field and linear terms in the vector
field $\omega$ as well as in the isovector $\rho$ and $\delta$
fields. An alternative would be to perform such a parametrisation of
the DBHF results in terms of density dependent relativistic hadron
theory (DDRH)~\cite{fule95,typel99,lenske00,niksic02,ava04} 
where non-linearity's due to higher order density corrections are 
absorbed into density dependent meson-nucleon vertices at the level 
of the effective Lagrangian.  The reason why we have chosen the 
standard RMF Lagrangian (\ref{eq:1}) is twofold: this allows a 
well defined low-density expansion while the density dependence 
of effective meson-nucleon vertices in DDRH depends on the choice 
of an particular ansatz for these functionals. Secondly, 
the extraction of such coupling functions from the present 
DBHF self-energies~\cite{dal04} shows that the DBHF vector self-energy, 
except for the very low density regime, has a linear density 
dependence which can be expressed by a linear  $\omega$ meson field. 
Non-linearities in the scalar channel are absorbed in the non-linear
$\sigma$ terms. 
The isovector dependence can also reasonably well be fitted 
through the two isovector   $\rho$ and $\delta$ mesons. 
In summary, such a procedure allows a well defined 
comparison of the microscopic DBHF model to RMF phenomenology and 
a controlled investigation of the low density regime, where the 
RMF fits break down and require additional correction terms, as 
will be seen in the following.

The adjusting procedure is twofold: firstly, we fit the parameters of
the RMF Lagrangian using the relativistic Dirac mass and the energy
density in symmetric and asymmetric nuclear matter obtained from the
DBHF calculation.
The fit includes 23 calculated points between $\rho_B$=0.02 and
0.13~fm$^{-3}$, plus two densities, $\rho_B$=0.1658~fm$^{-3}$ and
$\rho_B$=0.197~fm$^{-3}$, for $y=\rho_p/\rho_B$=0 to 0.5 with a 
step=0.05. 
We obtain the set of parameters called RMF presented in
Tab.~\ref{table:1}, by fitting the Dirac mass and the density of  
energy and imposing that the functional passes exactly through the
point at $\rho_B=0.197$ fm$^{-3}$ in symmetric nuclear matter and pure
neutron matter. 
The latter condition is imposed in order to get a value of 
the symmetry energy close to DBHF, as shown in Tab.~\ref{table:5}.
Then, we extract and fit the residual difference between the DBHF
calculation and the energy per particle in symmetric nuclear matter
and pure neutron matter separately. 
In order to check the sensitivity of the results on the functional
correction, we have investigated two different functionals.
In the following, we give the details of the adjusting procedure.

\begin{table}[htb] 
\centering
\begin{tabular}{|c|c|c|c|c|c|} 
\hline 
fits & $f_\sigma$ & $f_\sigma^{nl}$ & $f_\delta$ &
$f_\omega$ & $f_\rho$ \\
& [MeV$^{-1}$] & [MeV$^{-2}$fm$^3$] & [MeV$^{-1}$] &
[MeV$^{-1}$] & [MeV$^{-1}$]  \\ 
\hline  
RMF & 1.693 $10^{-2}$ & 3.735 $10^{-4}$ & 7.242
$10^{-3}$ & 1.299 $10^{-2}$ & 
8.843 $10^{-3}$ \\ 
set A NL$\rho\delta$~\cite{liu05} & 1.629 $10^{-2}$ & 9.35 $10^{-4}$ &
8.013 $10^{-3}$ & 1.18 $10^{-2}$ & 8.996 $10^{-3}$ \\ 
\hline 
\end{tabular} 
\caption{The parameters of the functional reproducing the scalar
mass and the energy density of the DBHF calculation is
compared to the parameters proposed in Ref~\cite{liu05}.}  
\label{table:1} 
\end{table}

\subsection{Determination of the $\sigma$ and $\delta$ coupling
constants} 

\begin{figure}[htb] 
\center 
\includegraphics[scale=0.3]{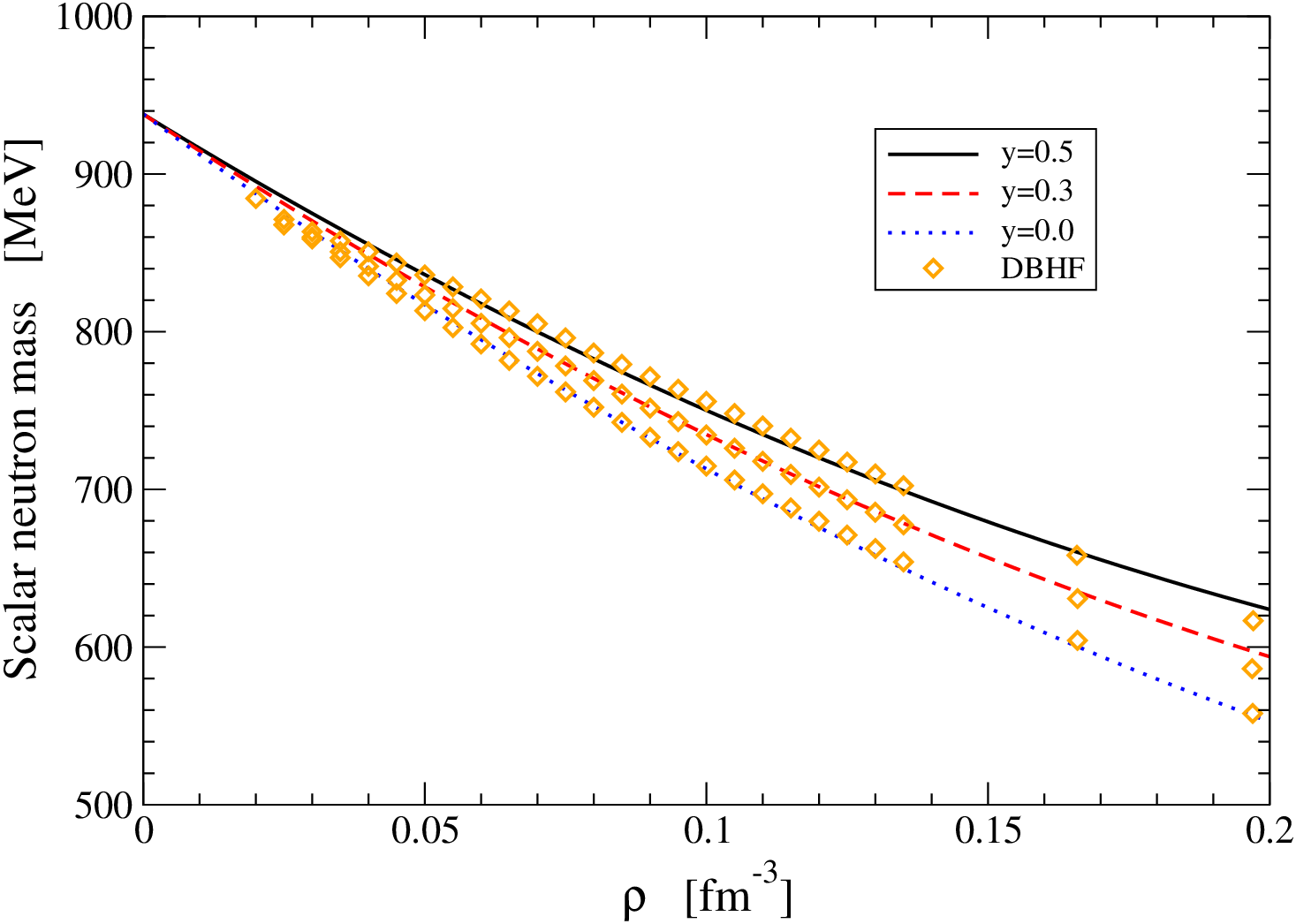} 
\caption{(Color online) Comparison between the best RMF adjustment
  and the scalar mass calculated with DBHF approach.
The asymmetry parameter $y=\rho_p/\rho_B$, ranges from 0.5 to 0.
It clearly shows a linear ($f_\sigma^2\rho_B$) and a quadratic
($f_\sigma^{nl}\rho_B^2$)  behavior in the baryonic density. 
The isospin asymmetry is also well reproduced by the linear term
  $f_\delta^2\rho_3$ in the asymmetry density $\rho_3$. } 
\label{figu01} 
\end{figure}

The low density expansion of the Dirac mass, Eq.~(\ref{eq:effmass}),
is used to determine the linear sigma coupling constant $f_\sigma$,
the scalar iso-vector $\delta$-meson $f_\delta$ and the non-linear
sigma coupling constant $f_\sigma^{nl}$. 
We deduce the value of the parameters for the adjustment to DBHF
results in asymmetric nuclear matter with $y=\rho_p/\rho_B$=0.5, 0.3
and 0.0. 
In Fig.~\ref{figu01} we compare the DBHF results with the best fit.
The linear contribution comes from the term $f_\sigma^2\rho_B$ in
Eq.~(\ref{eq:effmass}), then 
comes the quadratic term $f_\sigma^{nl}\rho_B^2$, the isospin
asymmetry is essentially coming from the first term $f_\delta^2\rho_3$
the contributions of the other terms are negligible.
The parameters are given in Tab.~\ref{table:1} and are compared to the
set A NL$\rho\delta$ proposed in Ref~\cite{liu05}.
The parameters $f_\sigma$ and $f_\delta$ are very similar for the two
set of parameters while the parameters $f_\sigma^{nl}$ differ by a
factor 3. In our fit, $f_\sigma^{nl}$ is obtained from the quadratic
density dependence of the scalar mass while in the set A
NL$\rho\delta$, the non-linear $\sigma$ coupling is adjusted to reduce
the compressibility modulus.
As a consequence, we obtain a lower value for the parameter
$f_\sigma^{nl}$, and the compressibility modulus is larger than
expected.
Moreover, to obtain a good fit of the binding energy, the non-linear
term $\epsilon_{NL,9}$ defined in Eq.~(\ref{eq:nl}) has to be divided
by a factor 2. This may indicate that higher order non-linearities in  
the $\sigma$
field and probably also non-linear $\omega$ terms should be taken into
account to obtain a proper description of the DBHF equation of state  
beyond saturation
density. However, to keep the formalism as simple as possible we
stick to the strandard NL model and apply this phenomenological correction.
The saturation properties are shown in Tab.~\ref{table:5}.
This indicate that one cannot reproduce the scalar mass density
dependence and the saturation density with standard $\sigma$-non
linear RMF Lagrangian. 
This illustrate the convenience of using a density functional where
the effects of the meson couplings are tractable.
In our Lagrangian, the compressibility modulus will be lowered by the
correction terms induced by the physics beyond the mean field.

\subsection{Determination of the $\omega$ and $\rho$ coupling constants} 

\begin{figure}[htb] 
\center 
\includegraphics[scale=0.4]{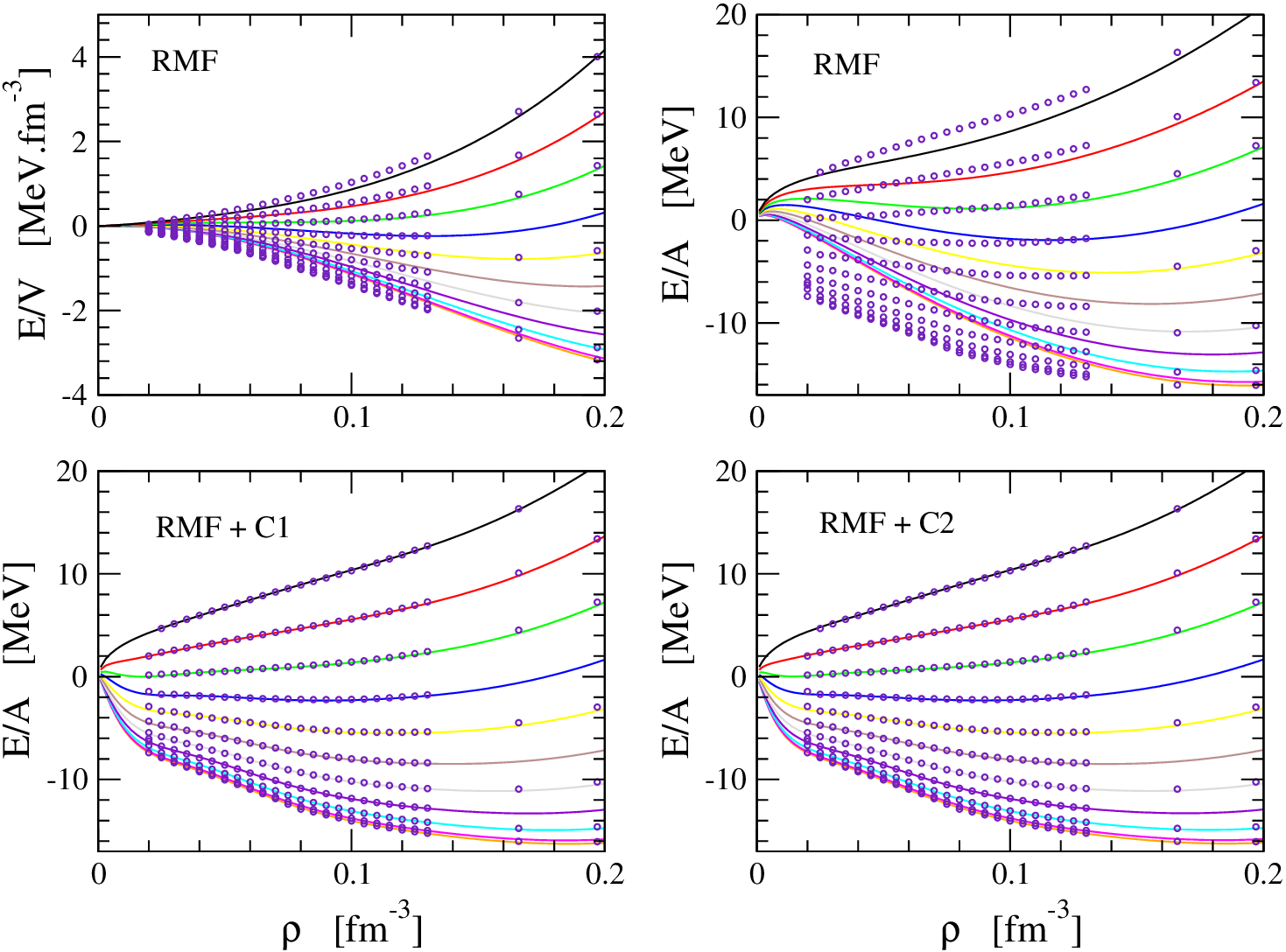}
\caption{(Color online) Comparison of the DBHF results (circles
  points) and  the low density functionals in asymmetric nuclear matter.
On the top panels are shown the energy density and the binding energy
  for the parameters RMF, while on the bottom panels are shown the
  binding energies for the corrected functionals RMF+C1 and RMF+C2.}
\label{artfig03} 
\end{figure} 

In contrast to DBHF theory, which shows a non-linear convergence to
zero in the binding energy, RMF theory converges smoothly.
We show in Fig.~\ref{artfig03} (top panel) a comparison between the 
DBHF calculation and the low density functional RMF for the density of 
energy, $E/V$, and the binding energy, $E/A$.
We remind that the non-linear term $\epsilon_{NL,9}$ defined in
Eq.~(\ref{eq:nl}) has to be divided by a factor 2.
The low density effects are reduced on the energy density plot
compared to the effects on the binding energy.
In fact, as $E/V=\rho_B E/A$, the low density behavior of the binding
energy is strongly reduced by the factor $\rho_B$ in the energy density.
Therefore, we adjust the mean field functional on the energy density
obtained from the DBHF calculation in asymmetric nuclear matter, where
the low density effects are weaker. We also force the functional 
to reproduce exactly the DBHF calculations at $\rho_B=0.197$ fm$^{-3}$
in symmetric nuclear matter and pure neutron matter. This constraint
ensure that the symmetry energy at saturation density is well
reproduced. 
The table~\ref{table:1} resumes the parameters obtained from the low
density functional RMF.
In the following section, we propose a correction of the functional
to take into account the correlations beyond the mean field.

\subsection{Corrections to the mean field functional RMF}

As already discussed and shown in Fig.~(\ref{fig01}) and 
Fig.~(\ref{artfig03}), ab-initio calculations like relativistic or 
non-relativistic Brueckner calculations have a completely different low 
density behavior compared to standard mean field prediction. 
For neutron matter this is a known fact and 
has e.g. also been investigated in the - hypothetical - unitary limit 
$a k_F \rightarrow \infty$ where $a$ is the $nn$ scattering length. In 
this limit many-body calculations 
(BHF, variational and GFMC) lead to a different low density behavior 
than RMF theory, see discussion in Ref. \cite{carlson03}. 
To account for the low density behavior of the DBHF equation of state
one has therefore to go beyond the standard prediction of mean field
theory.

In this paper we choose a pragmatic approach and propose a fit of the
difference between the DBHF equation of state and the low density RMF 
functional.
Thus we add two new functions $g^S(\rho_B)$ and $g^N(\rho_B)$ to the
energy density so that 
\begin{equation} 
\epsilon_{\rm DBHF} (\rho_n,\rho_p)=\epsilon_{\rm RMF}(\rho_n,\rho_p)
+(1-\beta^2)g^S(\rho_B)+\beta^2g^N(\rho_B) \; ,
\end{equation} 
where $\beta=(\rho_n-\rho_p)/\rho_B$. The additional terms $g^S(\rho_B)$ 
and $g^N(\rho_B)$ are respectively adjusted in symmetric nuclear matter 
and pure neutron matter. 
The isospin degree of freedom is factorized with a quadratic function
which respects the nuclear isospin symmetry. 
This approximation is often performed
(see for instance~\cite{bal04}) but in our case, it is also justified
afterwards by comparing the new functionals RMF+C1 and RMF+C2 to the
DBHF binding energies at low densities (Fig.~(\ref{artfig03}), bottom
panels). 
The functional correction is unknown, but it is clear that this
correction should be small around saturation density, and should
converge to zero at very small densities. 
Then an overall exponential shape
impose to fulfill the first condition and a factorization in power of the
density ensure that the second condition is also satisfied. 
We obtained two different density functionals which reproduce the data
with an equal accuracy.
The first phenomenological correction C1 is a product of a polynomial
function in the baryonic density with an exponential and has the
following form: 
\begin{eqnarray} 
g^S(\rho_B)&=& \frac{\rho_B}{0.06} \left( v^S_0+v^S_1\rho_B+v^S_2\rho_B^2\right) 
e^{-(\rho_B/\rho^S_0)^{1.2}} \; , \\
g^N(\rho_B)&=&  \left(\frac{\rho_B}{0.1}\right)^2 v^N_0 e^{-(\rho_B/\rho^N_0)^2}
\; , 
\end{eqnarray}
with the parameters:
$v^S_0$=-48.834 MeV.fm$^3$,
$v^S_1$=1073.3 MeV.fm$^6$,
$v^S_2$=-14813 MeV.fm$^9$,
$\rho^S_0$=0.03114 fm$^{-3}$,
$v^N_0$=5.373 MeV.fm$^6$,
$\rho^N_0$=0.0937 fm$^{-3}$.
The second phenomenological correction C2 is a sum of two exponentials
of the form: 
\begin{eqnarray} 
g^S(\rho_B)&=& v^S_0 \frac{\rho_B}{0.01} e^{-\rho_B/\rho^S_0}
+v^S_1\frac{\rho_B}{0.06}e^{-(\rho_B/\rho^S_1)^2}\; , \\
g^N(\rho_B)&=& v^N_0\frac{\rho_B}{0.01} e^{-(\rho_B/\rho^N_0)^{2.2}}
+v^N_1\frac{\rho_B}{0.1}e^{-(\rho_B/\rho^N_1)^2}\; , 
\end{eqnarray}
with the parameters:
$v^S_0$=-9.28 MeV.fm$^3$,
$v^S_1$=-5.48 MeV.fm$^3$,
$\rho^S_0$=0.0140 fm$^{-3}$,
$\rho^S_1$=0.0879 fm$^{-3}$,
$v^N_0$=-0.334 MeV.fm$^3$,
$v^N_1$=4.818 MeV.fm$^3$,
$\rho^N_0$=0.0629 fm$^{-3}$,
$\rho^N_1$=0.1046 fm$^{-3}$.
We expect to have a measure of the error induced by the peculiar
choice of the functional by comparing the prediction obtained with the
two functionals RMF+C1 and RMF+C2.

\begin{figure}[htb] 
\center 
\includegraphics[scale=0.4]{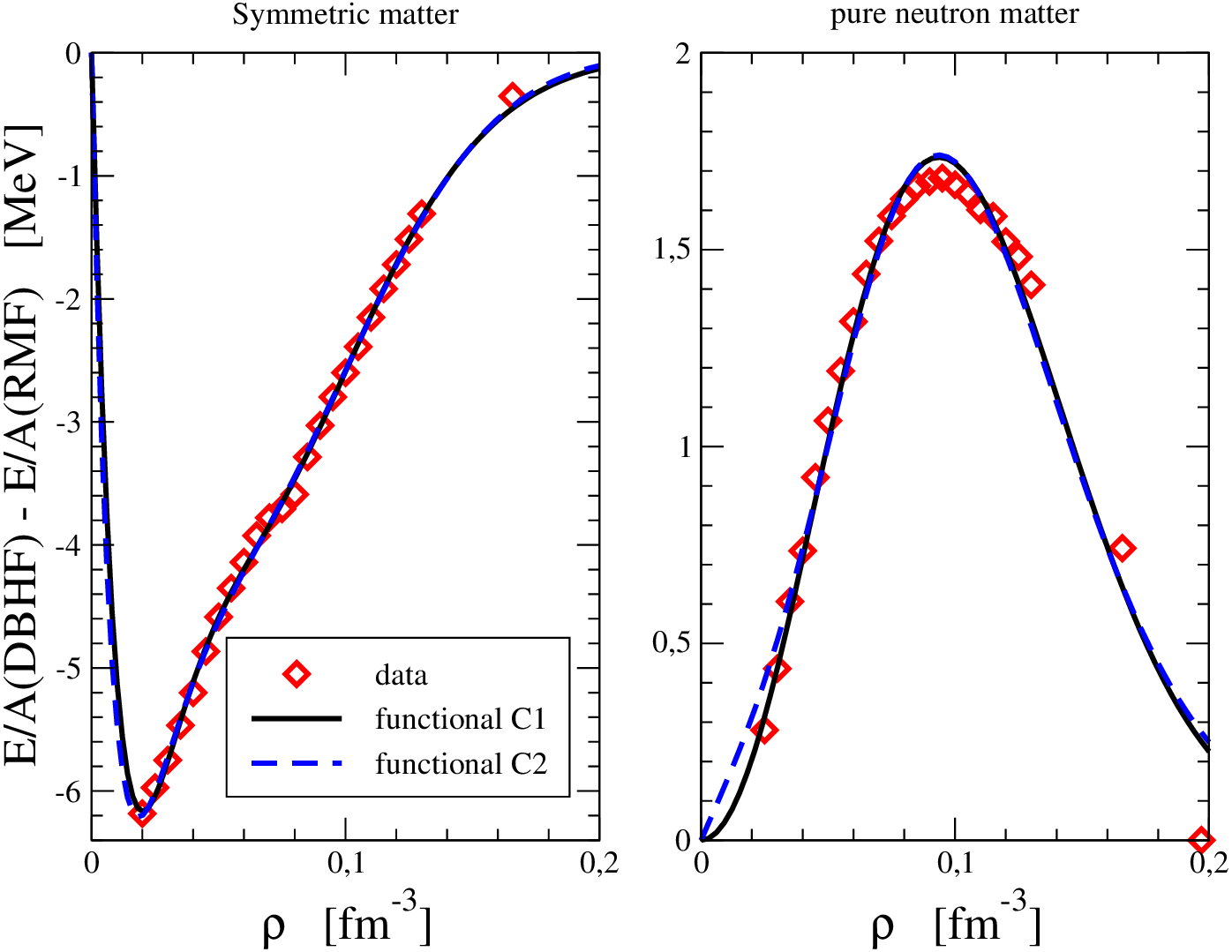}    
\caption{(Color online) Difference between the DBHF calculation and
  the low density RMF fit 
(square symbols). The corrections (solid lines) are drawn for symmetric
nuclear matter (left panel) and pure neutron matter (right panel).}  
\label{artfig02} 
\end{figure} 

In fig.~\ref{artfig02}, we shown the result of the adjustment of the 
functional C1 (solid line) and C2 (dashed line) to the difference
between the DBHF binding energy and the low density RMF functional
(square symbols) in symmetric nuclear matter (left panel) and pure
neutron matter (right matter).
Despite the different density dependence between the functionals C1
and C2, those two functionals reproduce the square symbols at the same 
level of accuracy.

\subsection{Properties of the functionals}

The table~\ref{table:5} give the properties of the functionals RMF, RMF+C1
and RMF+C2 around saturation density: the binding energy $B_0$, the saturation density 
$\rho_0$, the incompressibility $K_0$ and the symmetry energies $a_s^1$
and $a_s^2$. The properties of the DBHF calculation are also indicated. 
The properties of the low density RMF functional differ significantly
from the DBHF results. Indeed, the saturation density and the
compression modulus are higher than the DBHF results. 
It is a consequence of the low value of the parameters
$f_\sigma^{nl}$, as expected.
This parameter could not be changed, as it is already adjusted to the
quadratic density dependence of the scalar mass, 
However, the saturation properties of the corrected functionals RMF+C1 
and RMF+C2 are very close to the DBHF calculation.
Then, in our framework, the reduction of hte incompressibility modulus
is induced by the low density physics. This is a very different
understanding from the standard one which relies on non-linear
corrections at high densities.
We have also calculated the symmetry energy in two different manner:
either assuming a quadratic dependence in the asymmetry parameter $\beta$
($a_s^1$), or as we have a functional, by performing the second 
derivative around symmetric nuclear matter ($a_s^2$). 
The latest calculation is the exact one.
We note a systematic underestimation by about 1-2 MeV of the symmetry
energy assuming a quadratic behavior. 
This is a small error with respect to the difference in energy between
symmetric nuclear matter and pure neutron matter.

\begin{table}[htb] 
\begin{tabular}{|c|c|c|c|c|c|} 
\hline 
fits & $B_0 [MeV]$ & $\rho_0 [fm^{-3}]$ & $K_0 [MeV]$ & $a_s^1
[MeV]$ & $a_s^2 [MeV]$\\ 
\hline  
RMF    & -16.08 & 0.1933  & 365  & 35.8 & 37.7  \\ 
RMF+C1 & -16.27 & 0.1857  & 251  & 35.1 & 36.9  \\ 
RMF+C2 & -16.24 & 0.1856  & 242  & 35.1 & 36.9  \\ 
\hline 
DBHF & -16.15 & 0.1814  & 230  & 34.4 & - \\
\hline 
\end{tabular} 
\caption{Properties of the functionals. The symmetry energy has been
  obtain in two different ways: assuming a quadratic dependence of the
equation of state between symmetric and pure neutron matter ($a_s^1$),
or performing numerical derivation of the binding energy around
symmetric nuclear matter ($a_s^2$). A systematic difference is
observed but is less than 5\%.}
\label{table:5} 
\end{table} 
 
In this context it is worth to mention that RMF fits to finite 
nuclei require relatively high compression moduli $K\sim 300$ 
MeV~\cite{lal97,ring96}. Equations of state with a stiff high density 
behavior stand, however, in contrast to the information extracted 
from heavy ion reactions~\cite{sturm01,fuchs01}. 
The pure RMF fits to the DBHF EOS, i.e. discarding the low density
correction term, provide equations of state which are stiff, however,
not due to their high density behavior but do to the low density
part. The compression moduli of the pure RMF contributions without
correction term are K=365~MeV in contrast to the soft DBHF EOS with
K=230~MeV.  
If one assumes that the correction terms contain effectively
contributions from the deuteron and/or reflect the precursor of a
superfluid low density state which leads to additional binding in
infinite matter, but plays no substantial role in finite nuclei, this
could explain the discrepancy between the EOS obtained from RMF fits
to finite nuclei and that predicted by DBHF or the a low density
virial expansion~\cite{horowitz05}.

\section{Spinodal Instabilities} 
\label{sec:si}
 
Let us consider asymmetric nuclear matter characterized by a proton
and a neutron densities $\rho _{i}=$ $\rho _{p}$, $\rho _{n}$. 
In infinite matter, the extensively of the free energy
implies that it can be reduced to a free energy density~: $F(T,V,N_{i})=V%
\mathcal{F}(T,\rho _{i}).$ The system is stable against separation
into two phases if the free energy of a single phase is lower than the
free energy in all two-phases configurations. 
This stability criterion implies that the free
energy density is a convex function of the densities $\rho _{i}$. A local
necessary condition 
is the positivity of the curvature matrix~: 
\begin{equation}
\left[ \mathcal{F}_{ij}\right] =\left[ \frac{\partial ^{2}\mathcal{F}}{%
\partial \rho _{i}\partial \rho _{j}}|_{T}\right] \equiv \left[ \frac{%
\partial \mu _{i}}{\partial \rho _{j}}|_{T}\right]  \label{eq5}
\end{equation}
where we have introduced the chemical potentials $\mu _{j}\equiv \frac{%
\partial F}{\partial N_{j}}|_{T,V,N_{i}}=\frac{\partial \mathcal{F}}{%
\partial \rho _{j}}|_{T,\rho _{i\not{=}j}}$. In the considered two-fluids
system, the $[\mathcal{F}_{ij}]$ is a $2*2$ symmetric matrix, so it has 2
real eigenvalues $\lambda ^{\pm}$~: 
\begin{equation}
\lambda ^{\pm}=\frac{1}{2}\left( \mathrm{Tr}\left[ \mathcal{F}_{ij}\right]
\pm \sqrt{\mathrm{Tr}\left[ \mathcal{F}_{ij}\right] ^{2}-4\mathrm{Det}\left[ 
\mathcal{F}_{ij}\right] }\right)  \label{eq23}
\end{equation}
associated to 
eigenvectors $\mathbf{\delta \rho }^{\pm}$ defined by ($i\neq j$) 
\begin{equation}
\frac{{\delta \rho }_{j}^{\pm}}{{\delta \rho }_{i}^{\pm}}=\frac{\mathcal{F}%
_{ij}}{\lambda ^{\pm}-\mathcal{F}_{jj}}=\frac{\lambda ^{\pm}-\mathcal{F}_{ii}%
}{\mathcal{F}_{ij}}  \label{eq25}
\end{equation}
Eigenvectors associated with negative eigenvalue indicate the
direction of the instability. It defines a local order parameter since
it is the direction along which the phase separation occurs. 
The eigen values $\lambda $ define sound velocities, $c$, by 
${c}^{2}=\frac{1}{18m}\rho _{1}\,\lambda$. 
In the spinodal area, the eigen value $\lambda$
is negative, so the sound velocity, $c$, is purely imaginary and the
instability time $\tau $ is given by $\tau =d/|c|$ where $d$ is a typical
size of the density fluctuation.

The requirement that the local curvature is positive 
is equivalent to the requirement that both the trace 
($\mathrm{Tr}[\mathcal{F}_{ij}]=\lambda ^{+}+\lambda ^{-}$) 
and the determinant 
($\mathrm{Det}[\mathcal{F}_{ij}]=\lambda ^{+}\lambda ^{-}$) are positive 
\begin{equation}
\mathrm{Tr}[\mathcal{F}_{ij}]\geq 0,\hbox{ and }\mathrm{Det}[\mathcal{F}%
_{ij}]\geq 0  \label{eq6}
\end{equation}
The use of the trace and the determinant which are two basis-independent
characteristics of the curvature matrix clearly stresses the fact that
the stability analysis should be independent of the arbitrary choice
of the thermodynamical quantities used to label the state e.g. 
$(\rho _{p}$, $\rho_{n})$ or $(\rho _{1}$, $\rho _{3})$. 
If Eq.~(\ref{eq6}) is violated the system is in the unstable region of
a phase transition.

\begin{figure}[htb] 
\center 
\includegraphics[scale=0.5]{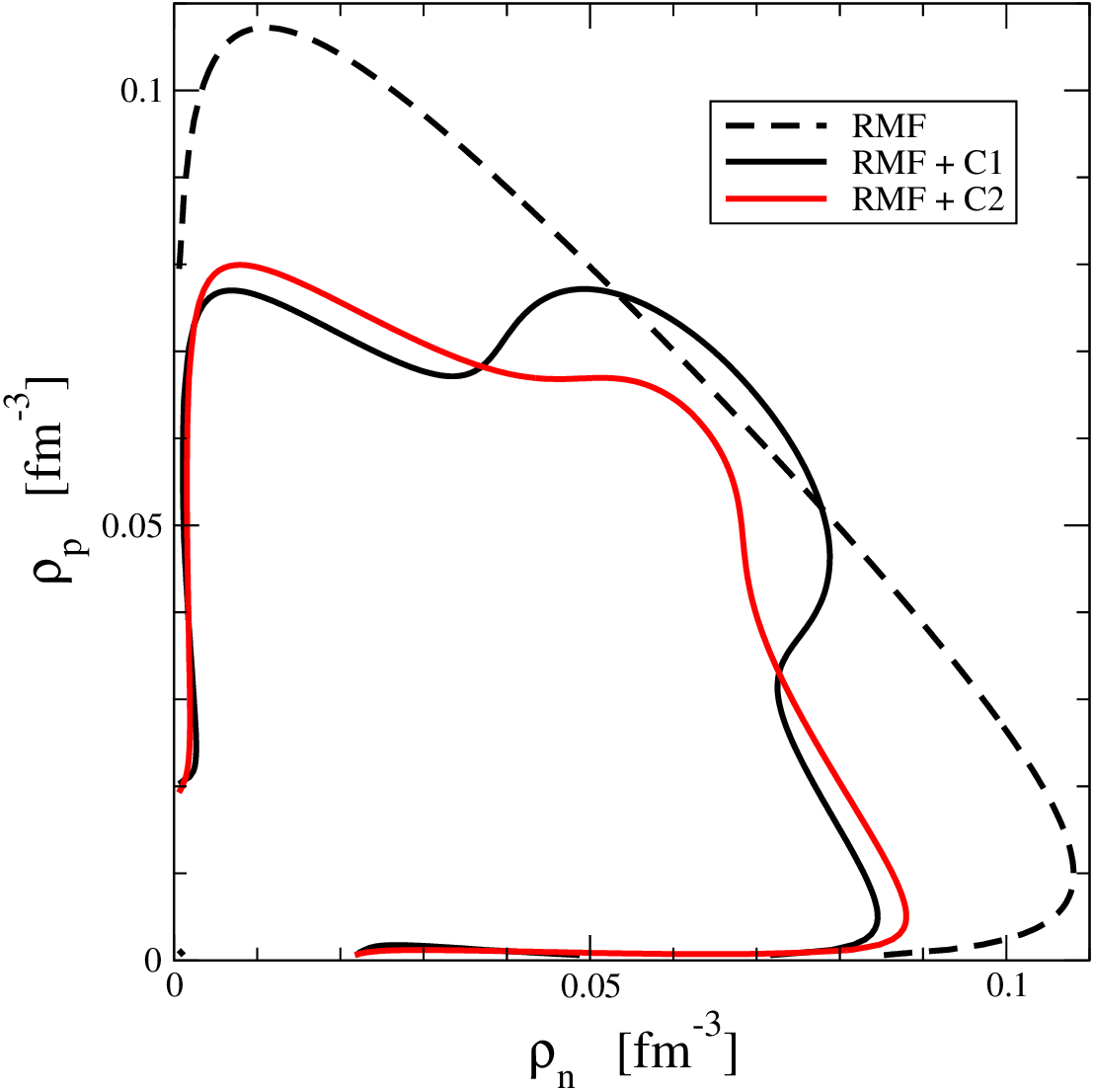} 
\caption{(Color online) Spinodal contour for the low density RMF
  functional, RMF, RMF+C1 and RMF+C2.} 
\label{artfig06} 
\end{figure}

We represent in Fig.~(\ref{artfig06}) the spinodal contour for the low 
density RMF functional, RMF, RMF+C1 and RMF+C2. The contour for the RMF 
functional is very similar to the one obtained previously with 
non-relativistic interactions~\cite{bar06,duc06,duc07}. 
Meanwhile, spinodal densities are a little bit larger. Indeed, in
mean-field models, it is well known that the spinodal density in
symmetric nuclear matter scales with the saturation density~\cite{cho04}.
The spinodal densities we obtain is just due to a scaling effect
induced by the saturation density which is slightly to large for the
functional RMF.
The corrections induce important modifications of the spinodal
contour, especially in asymmetric nuclear matter. 
For an asymmetry parameter of about $y\sim 0.4$, the spinodal density,
$\rho_s$, is reduced by about 15-20\%.

\begin{figure}[htb] 
\center 
\includegraphics[scale=0.5]{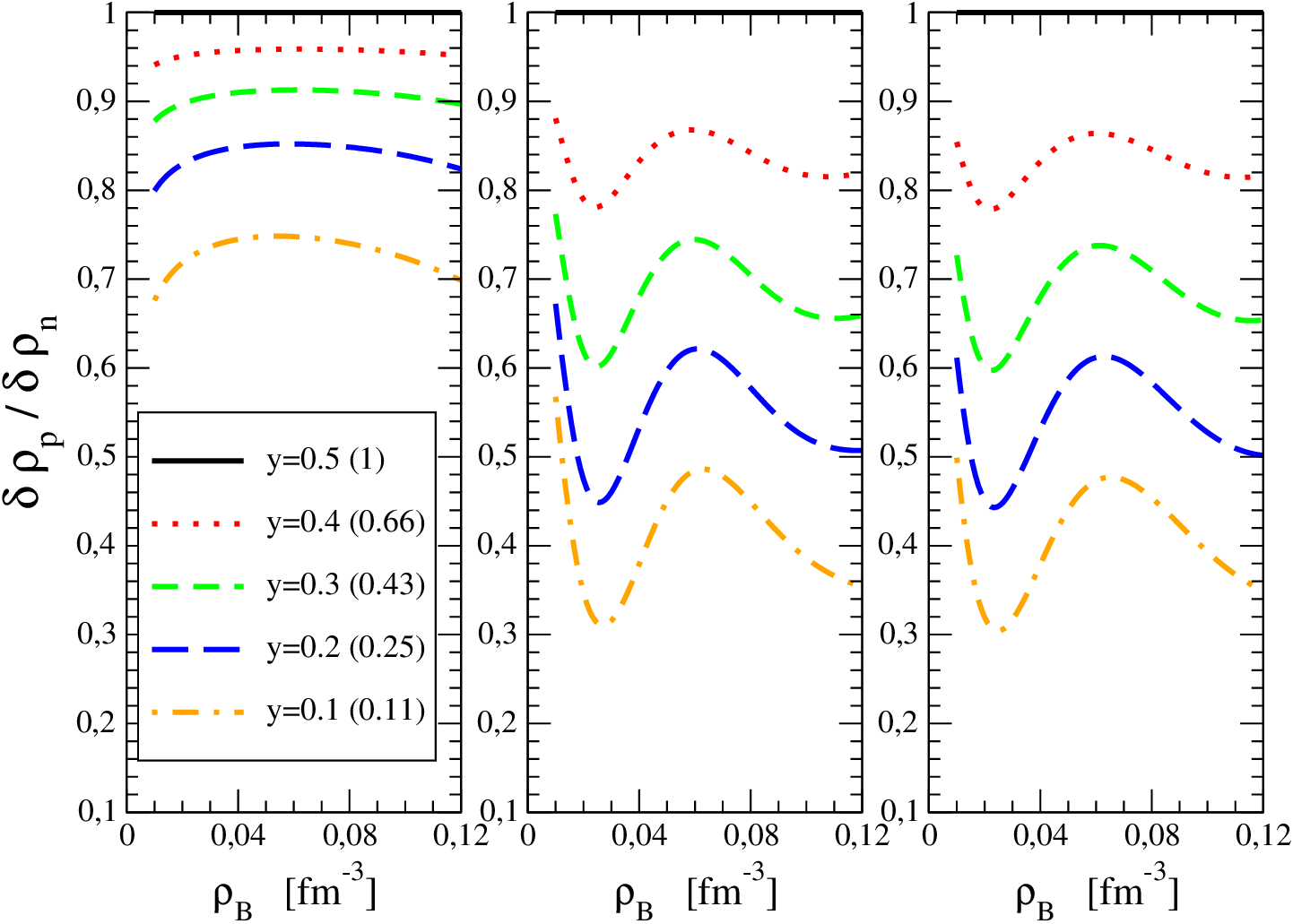} 
\caption{(Color online) Evolution of the unstable mode
  $\delta\rho_n/\delta\rho_p$ 
with the density for asymmetries ranging from $y$=0.1 to 0.5.}
\label{artfig07}
\end{figure}

In usual mean field calculations, it has been found that the direction
of the unstable mode is still isoscalar in asymmetric nuclear
matter~\cite{bar06}.
As the spinodal contour calculated with the functionals RMF+C1 and RMF+C2
differs significantly from the functional RMF, one could wonder what
would be the consequences for the unstable mode: still isoscalar or
possibly isovector ?
We have represented our results on Fig.~(\ref{artfig07}) where is
shown the evolution of the unstable mode $\delta\rho_n/\delta\rho_p$
with the density for asymmetries ranging from $y$=0.1 to 0.5. We
represent only neutron rich matter, proton rich is easily deduced
from the isospin symmetry property.
An isoscalar mode is defined by $\delta\rho_n/\delta\rho_p$=1 while a
mode along $y$=cst satisfies $\delta\rho_n/\delta\rho_p=y/(1-y)$. 
For the convenience of understanding Fig.~(\ref{artfig07}), the value
$y/(1-y)$ is written into parenthesis in the legend of each curve.
For the three functionals (RMF, RMF+C1 and RMF+C2) the unstable mode is
included in between the isoscalar direction and the direction $y$=cst. 
However, the results obtained with the functionals RMF+C1 and RMF+C2
are very similar and differ from the one obtained with the functional
RMF.  
Indeed, for the functionals RMF+C1 and RMF+C2, the unstable mode is
less isoscalar than the one calculated with the functional RMF, hence,
the fractionation mechanism should be less pronounced than the one
predicted with mean-field models~\cite{bar06,duc06,duc07}.
The gas phase should then be less asymmetric than what was
previously predicted based on mean-field calculations.

\section{Conclusions} 
\label{sec:c}

We have obtained two functionals which give a very good description of
the DBHF calculations in asymmetric nuclear matter. 
Those functionals are based on the low density RMF Lagrangian which is 
developed as a series expansion of $k_F/M$. Furthermore effective
terms are added which account for effects beyond mean field like the
deuteron pole.  
This parametrisation has been used to understand the topological
properties of the energy density, like the spinodal zone. We have
observed that the spinodal zone is reduced in asymmetric nuclear matter 
contrarily to all the previous mean-field calculation. 
Based on the analysis of the direction of the unstable mode, it is shown
that the gas phase could be less asymmetric than what was previously 
predicted based on mean-field calculations.

This calculation has been performed at zero temperature while
experiments probe the liquid-gas phase transition near the critical 
temperature~\cite{cho04}.
An extension of this work to finite temperature is then necessary, but
one could expect from mean field calculations that the critical
density scales with the spinodal density $\sim 0.5\rho_s$.
In a future work, it would then be interesting to evaluate the effects
of the two-body correlations at finite temperature.

As a final conclusion and outlook of this work, we would like to
stress that the low density EoS is required for both heavy ion
collisions as well as for the description of the crust of neutron
stars where a low density neutron gas is formed. 
The density functional theory is then an interesting framework where
these two different nuclear systems could be describe by a unique
functional.

\begin{acknowledgements}
This work has been supported by the Deutsche Forschungsgemeinschaft (DFG) under contract no. FA 67/29-1 and by the Spanish Ministry of
Education and Science under grant no. SB-2005-0131.
\end{acknowledgements}

\appendix

\section{Relativistic Mean Field Model} 
\label{app:rmfm}

A set of coupled field equations for the meson and nucleon fields can
be obtained from the Lagrangian in Eq.~(\ref{eq:1}). This field
equations in a mean field approximation ($MFA$) are   
\begin{eqnarray}\label{eq:2} 
&& (i\gamma_{\mu}\partial^{\mu}-(M- g_{\sigma}\sigma 
-g_\delta{\tau_3}\delta_3)-g_\omega\gamma^{0}{\omega_0}-
g_\rho\gamma^{0}{\tau_3}{\rho_0})\psi=0,\\ \label{eq:2b}
&& m_{\sigma}^2\sigma+ a{{\sigma}^2}+ b{{\sigma}^3}=
g_\sigma<\bar\psi\psi>=g_\sigma{\rho}_s,\\ \label{eq:2c}
&& m^2_{\omega}\omega_{0}=g_\omega<\bar\psi{\gamma^0}\psi>=
g_\omega\rho_B,\\ \label{eq:2d}
&& m^2_{\rho}\rho_{0}=g_\rho<\bar\psi{\gamma^0}\tau_3\psi>=
g_\rho\rho_3,\\  \label{eq:2e}
&& m^2_{\delta}\delta_3=g_{\delta}<\bar\psi\tau_3\psi>=g_{\delta}\rho_{s3}, 
\end{eqnarray} 
where $\rho_3=\rho_p-\rho_n$ and $\rho_{s3}=\rho_{sp}-\rho_{sn}$,  
$\rho_B=\rho_p+\rho_n$ and  
\begin{eqnarray} 
\rho_{si}=\frac{\gamma}{(2\pi)^3}\int_0^{k_{F_i}} d^3k \frac{M^*_i}{E^*_i(k)} 
\end{eqnarray} 
are the baryon and the scalar densities, respectively. 
 
Neglecting the derivatives of mesons fields, the energy-momentum tensor in 
$MFA$ is given by 
\begin{equation}\label{eq:3} 
T_{\mu\nu}=i\bar{\psi}\gamma_{\mu}\partial_{\nu}\psi+[\frac{1}{2} 
 m_{\sigma}^2\sigma^2+U(\sigma)+\frac{1}{2}m_{\delta}^2\vec{\delta^2} 
-\frac{1}{2}m^2_{\omega}\omega_{\lambda} \omega^{\lambda} 
-\frac{1}{2}m^2_{\rho}\vec{\rho_{\lambda}}\vec{\rho}^{\lambda}]g_{\mu\nu}~. 
\end{equation}

The equation of state ($EOS$) for nuclear matter at T=0 is straightforwardly 
obtained from the energy-momentum tensor. 
The energy density has the form 
\begin{equation}\label{energy2} 
\epsilon=<T^{00}>= 
\sum_{i=n,p}{2}\int \frac{{\rm d}^3k}{(2\pi)^3}E_{i}^*(k) 
+\frac{1}{2}m_\sigma^2\sigma^2+U(\sigma) 
+\frac{1}{2}m_\omega^2\omega_0^2 
+\frac{1}{2}m_{\rho}^2 \rho_0^2, 
+ \frac{1}{2}m_{\delta}^2\delta_3^2 ~.
\end{equation} 
The pressure is given by 
\begin{equation}\label{eq:5} 
p =\frac{1}{3}\sum_{i=1}<T^{ii}>= \sum_{i=n,p}\frac{2}{3}
\int \frac{{\rm d}^3k}{(2\pi)^3} \frac{k^2}{E_{i}^*(k)}
 -\frac{1}{2}m_\sigma^2\sigma^2-U(\sigma)+ 
\frac{1}{2}m_\omega^2\omega_0^2 
+\frac{1}{2}m_{\rho}^2{\rho_0^2} 
-\frac{1}{2}m_{\delta}^2\delta_3^2~, 
\end{equation} 
where $E_i^*(k)=\sqrt{k^2+{{M_i}^*}^2}$, $i=p,n$. The nucleon 
Dirac masses are, respectively 
\begin{eqnarray}
{M_p}^*=M-g_\sigma\sigma-g_\delta\delta_3~, \label{eq:6} \\
{M_n}^*=M-g_\sigma\sigma+g_\delta\delta_3~. \label{eq:7} 
\end{eqnarray} 
In mean field approximation the kinetic contributions to energy density and 
pressure in Eqs. (\ref{energy2}) and (\ref{eq:5}) can 
easily be evaluated by partial integration 
which yields 
\begin{eqnarray} 
\epsilon_{kin} &=& 
\sum_{i=n,p}{2}\int \frac{{\rm d}^3k}{(2\pi)^3}E_{i}^*(k) =
\sum_{i=n,p}\left( \frac{3}{4}\rho_iE^*_i(k_{F_i})
+\frac{1}{4}M^*_i\rho_{si} \right) \\
p_{kin} &=& \sum_{i=n,p}\frac{2}{3}
\int \frac{{\rm d}^3k}{(2\pi)^3} \frac{k^2}{E_{i}^*(k)} 
= \sum_{i=n,p}\left(  \frac{1}{4}\rho_iE^*_i(k_{F_i})-
\frac{1}{4}M^*_i\rho_{si}\right)   ~.
\end{eqnarray} 

The nucleon chemical potentials $\mu_i$ are given in terms of the 
vector meson mean fields 
\begin{equation}\label{eq:8} 
\mu_i=\sqrt{k_{F_i}^2+{{M_i}^*}^2}+ g_\omega\omega_0\mp g_{\rho}\rho_0\ 
~~(+~{\rm proton}, -~{\rm neutron})~, 
\end{equation} 
 where the proton/neutron Fermi momenta $k_{F_i}$ are related to the 
corresponding densities by  $k_{F_i}=(3\pi^2 \rho_i)^{1/3}$~.

\section{Solution method of the non-linear self-consistent equation
for the isoscalar scalar field $\sigma$.}
\label{app:nls}
Our starting point is the self-consistent equation in (\ref{scalarself}).
Let's call $\sigma_0$ the solution of this linear self-consistent equation,  
$g_\sigma \sigma_0=f_\sigma^2\rho_s$, and $\sigma_1$ the first order
correction induced by the non-linear terms. Then $\sigma_1$ fulfils
the following equation 
\begin{eqnarray} 
g_\sigma \sigma_1 = - \frac{a}{g_\sigma m_\sigma^2} 
(g_\sigma\sigma_0+g_\sigma\sigma_1)^2  
- \frac{b}{(m_\sigma g_\sigma)^2}(g_\sigma\sigma_0+g_\sigma\sigma_1)^3 ~. 
\label{self1} 
\end{eqnarray} 
We first suppose that $\sigma_1/\sigma_0\ll 1$ (we will verify this hypothesis 
after-while).  
Then, Eq.~(\ref{self1}) leads to 
\begin{eqnarray} 
\frac{\sigma_1}{\sigma_0} = -\frac{a}{m_\sigma^2}g_\sigma \sigma_0 
\left(1+2\frac{\sigma_0}{\sigma_1}\right) 
-\frac{b}{g_\sigma m_\sigma^2}\left(g_\sigma \sigma_0\right)^2 
\left(1+3\frac{\sigma_0}{\sigma_1}\right) 
+ { o}(\frac{\sigma_1}{\sigma_0}^2)~. 
\end{eqnarray}  

Indeed, it reads 
$2a\frac{\hbar}{m_\sigma}x+\frac{3b}{g_\sigma}x^2$ where 
$x=g_\sigma\sigma_0/m_\sigma=g_\sigma^2\left(\frac{\hbar}{m_\sigma}\right)^3\rho_s$. 
With typical values e.g. from the NL3 model, i.e. 
$g_\sigma\sim 10$, $m_\sigma\sim 500$ MeV, 
the parameter $x$ is approximately $x=1.6$ at saturation density. 
Then, the two terms of the denominator are about 20. 
The correction $\sigma_1/\sigma_0$ can be expressed as a function of the 
scalar density as 
\begin{eqnarray} 
\frac{\sigma_1}{\sigma_0} = \frac 
{-\frac{a g_\sigma^2}{m_\sigma^4}\rho_s 
-\frac{b g_\sigma^3}{ m_\sigma^6}\rho_s^2} 
{1+2\frac{a g_\sigma^2}{m_\sigma^4}\rho_s 
+3\frac{b g_\sigma^3}{ m_\sigma^6}\rho_s^2} 
+ { o}(\frac{\sigma_1}{\sigma_0}^2) 
\end{eqnarray} 
and taking into account only first order corrections, we arrive at 
\begin{eqnarray} 
g_\sigma \sigma_1 = - \frac{a}{g_\sigma m_\sigma^2} (g_\sigma\sigma_0)^2  
+\left(\frac{2a^2}{g_\sigma^2 m_\sigma^4}- \frac{b}{(m_\sigma 
  g_\sigma)^2}\right) (g_\sigma\sigma_0)^3 + { o}(g_\sigma\sigma_1^2) ~.
\end{eqnarray} 
We keep only the first term which contributes up to the power 9 in 
$k_{Fi}$. Hence, the approximate solution of the non-linear self-consistent 
equation is 
\begin{eqnarray} 
g_\sigma \sigma &=& f_\sigma^2 \rho_s - \frac{a f_\sigma^4}{g_\sigma m_\sigma^2}
\rho_s^2 + { o}(k_{Fi}^9). 
\label{scalarapp}
\end{eqnarray} 
The first term on the r.h.s. of Eq.~(\ref{scalarapp}) is the solution
of the linear self-consistent equation, while the second term is
induced by non-linear corrections.

\end{document}